\documentclass[12pt]{article}
\usepackage[english]{babel}
\usepackage{amsmath}
\usepackage{amssymb}
\usepackage{bbm}
\usepackage{color}
\usepackage{graphicx}
\usepackage{epsfig}
\usepackage[titletoc,toc,title]{appendix}
\Roman{section}
\newcommand{\naw}[1]{\left(#1\right)}
\newcommand{\ket}[1]{\left|#1\right>}
\newcommand{\bra}[1]{\left<#1\right|}
\newcommand{\av}[1]{\left<#1\right>}
\newcommand{\com}[1]{\left[#1\right]}
\newcommand{\modu}[1]{\left|#1\right|}
\newcommand{\poisson}[1]{\left\{#1\right\}}

\begin{document}

\begin{center}
\textsc{\Large{Breaking of Bell inequalities from $S_4$ symmetry: the three orbits case}}
\newline

\large{Katarzyna Bolonek-Laso\'n}\footnote{kbolonek@uni.lodz.pl}\\ 
\emph{\normalsize{Faculty of Economics and Sociology, Department of Statistical Methods, \\University of Lodz,
41/43 Rewolucji 1905 St., 90-214 Lodz,  Poland.}}\\
\large{\'Scib\'or  Sobieski}\footnote{scibor.sobieski@uni.lodz.pl}\\ 
\emph{\normalsize{Faculty of Physics and Applied Informatics, Department of Computer Science, \\University of Lodz,
149/153 Pomorska St., 90-236 Lodz,  Poland.}}\\

\end{center}
\begin{abstract}
The recently proposed (\emph{Phys. Rev.} \textbf{A90} (2014), 062121 and \emph{Phys. Rev.} \textbf{A91} (2015), 052110) group theoretical approach to the problem of breaking the Bell inequalities is applied to $S_4$ group. The Bell inequalities based on the choice of three orbits in the representation space corresponding to standard representation of $S_4$ are derived and their breaking is described. The corresponding nonlocal games are analyzed. 
\end{abstract}
\section{Introduction}
The famous Bell inequalities \cite{bell} provide the necessary conditions for any theory to be a local realistic one. Their importance stems from the observation  that they can be violated in quantum theory. As a result the  Bell inequalities can be used for test of entanglement and as a basis for protocols in quantum cryptography \cite{ekert}.

Bell inequalities have been studied intensively by numerous authors. Their various forms characterized by the number of parties, measurement  settings and outcomes for each measurement have been derived \cite{clauser}$\div$\cite{cabello} (for a review, see \cite{liang} and \cite{brunner}).

Recently, there appeared interesting papers \cite{ugur}, \cite{ugur1} where the group theoretical method have been proposed as a tool for analyzing the quantum mechanical violation of Bell inequalities. Examples of Bell inequalities based on representations of some finite groups were presented there. Further example has been considered in Ref. \cite{bolonek}. It is based on $S_4$ symmetry and its standard irreducible representation. The resulting Bell inequality is obtained by selecting two generic orbits determined by the geometry of tetrahedron. In the present paper we provide further examples of Bell inequalities related to the symmetry of tetrahedron. They result from the particular choices of three generic orbits. 

The paper is organized as follows. In Sec. II we discuss the Bell inequalities from the point of view of the existence of joint probability distribution and describe the group theoretical approach to the problem of their breaking proposed by G\H uney and Hillery. This approach is then applied in Sec. III to the symmetric group $S_4$. Three examples of quantum mechanical breakdown of Bell inequalities are presented. They are based on the specific choice of orbits in standard representation of $S_4$. In each of three cases we consider the set of states arising from the choice of three orbits. The results are interpreted in Sec. IV in the framework of game theory. Some technical detailes are relegated to the Appendix. 

\section{Bell inequalities}

Quantum  mechanical violation of Bell inequalities is closely related to the existence of noncommuting observables. In two elegant papers \cite{fine}, \cite{fine1} Fine provided a particulary transparent interpretation of Bell inequalities (see also \cite{halliwell}, \cite{halliwell1}). Assume we have a number of random variables possessing joint probability distribution. Bell inequalities concern the joint probability distributions of some subsets of the initial set of random variables. They result from the assumption that the latter can be obtained as marginals from the original joint probability distribution. What is even more important, the Bell inequalities form also the sufficient conditions for the existence of joint distribution returning other probabilities as marginals. In fact, the latter condition provides a set of linear equations for the joint distribution which possess the whole family of solutions. 
We are interested in solutions belonging to the interval $\av{0,1}$. The possibility of selecting such solutions relies on the validity of Bell inequalities. 

Fine's theorem explains the origin of quantum machanical violation of Bell inequalities. Due to the uncertainty principle the joint probability can be constructed only for the set of mutually commuting observables. Therefore, no inequality of Bell type could be derived for joint probabilities of commuting observables if they would follow from the assumption that these probabilities emerge as marginals from joint distribution for larger set of in general noncommuting observables.

Let us illustrate the above discussion by a simple example. Let $\hat{A}$  be some observable with the spectral decomposition
\begin{equation}
\hat{A}=\sum_{i}a_i\hat{\Pi}_i
\end{equation} 
where $\hat{\Pi}_i$ are the projectors on the relevant eigenspaces (we shall assume our space of states in finitedimensional). Consider any state $\hat{\rho}$ and let \cite{santos}
\begin{equation}
C\naw{\zeta}=\text{Tr}\naw{e^{i\zeta\hat{A}}\hat{\rho}}=\sum_i e^{i\zeta a_i}\text{Tr}\naw{\hat{\Pi}_i\hat{\rho}}\equiv\sum_i e^{i\zeta a_i}p_i
\end{equation}
be the generating function for the moments of $\hat{A}$:
\begin{equation}
\av{\hat{A}}=\text{Tr}\naw{\hat{A}^n\hat{\rho}}=\sum_i
 a_i^n p_i=\naw{-i\frac{d}{d\zeta}}^n C\naw{\zeta}\Big |_{\zeta=0}.
 \end{equation}
The probability distribution is obtained by Fourier transform
\begin{equation}
p\naw{a}=\frac{1}{2\pi}\int d\zeta e^{-i\zeta a}C(\zeta)=\sum_i p_i\delta\naw{a-a_i}.
\end{equation}
Assume now we have two observables,
\begin{equation}
\hat{A}_1\equiv \sum_i a_{1i}\hat{\Pi}_{1i},\qquad \hat{A}_2\equiv\sum_k a_{2k}\hat{\Pi}_{2k}.
\end{equation}
The generating function for the moments $\av{A_1^{n_1}A_2^{n_2}}$ reads
\begin{equation}
\begin{split}
& C\naw{\zeta_1,\zeta_2}=\text{Tr}\naw{e^{i\zeta_1\hat{A}_1}e^{i\zeta_2\hat{A}_2}\hat{\rho}}=\sum_{i,k}e^{i\zeta_1 a_{1i}}e^{i\zeta_2 a_{2k}}\text{Tr}\naw{\hat{\Pi}_{1i}\hat{\Pi}_{2k}\hat{\rho}}\equiv\\
&\qquad\qquad \equiv \sum_{i,k}e^{i\zeta_1 a_{1i}}e^{i\zeta_2 a_{2k}}p_{ik}.
\end{split}
\end{equation}
We are tempetd to define the joint probability as
\begin{equation}
\begin{split}
& p\naw{a_1,a_2}\equiv\frac{1}{4\pi^2}\int d\zeta_1 d\zeta_2 e^{-i\naw{\zeta_1a_1+\zeta_2a_2}}C\naw{\zeta_1,\zeta_2}=\\
&\qquad\qquad =\sum_{i,k}p_{ik}\delta\naw{a_1-a_1i}\delta\naw{a_2-a_{2k}}.\label{a1}
\end{split}
\end{equation}
Due to
\begin{equation}
\sum_{k}p_{ik}=\sum_k\text{Tr}\naw{\hat{\Pi}_{1i}\hat{\Pi}_{2k}\hat{\rho}}=\text{Tr}\naw{\hat{\Pi}_{1i}\naw{\sum_k\hat{\Pi}_{2k}}\hat{\rho}}=\text{Tr}\naw{\hat{\Pi}_{1i}\hat{\rho}}=p_{1i}
\end{equation}
single probability densities can be obtained as marginals
\begin{equation}
p_1\naw{a_1}=\int da_2 p\naw{a_1,a_2}.
\end{equation}
To have the genuine probability distribution we must assume $p_{ik}\geq 0$. Then the last expression (\ref{a1}) provides a finite positive measure on $\mathbb{R}^2$. Therefore, by Bochner theorem $C\naw{\zeta_1,\zeta_2}$ is positive definite function \cite{reed}. In particular
\begin{equation}
C\naw{\zeta_1,\zeta_2}=\overline{C\naw{-\zeta_1,-\zeta_2}}
\end{equation}
or
\begin{equation}
\text{Tr}\naw{e^{i\zeta_1\hat{A}_1}e^{i\zeta_2\hat{A}_2}\hat{\rho}}=\text{Tr}\naw{e^{i\zeta_2\hat{A}_2}e^{i\zeta_1\hat{A}_1}\hat{\rho}}.\label{a2}
\end{equation}
Assuming that (\ref{a2}) holds for all states $\hat{\rho}$ we find
\begin{equation}
e^{i\zeta_1\hat{A}_1}e^{i\zeta_2\hat{A}_2}=e^{i\zeta_2\hat{A}_2}e^{i\zeta_1\hat{A}_1}
\end{equation}
or $\com{\hat{A}_1,\hat{A}_2}=0$. We see that the joint probability can be defined only for commuting variables.

Taking into account Fine's results one concludes that the general scheme for deriving the Bell inequalities is quite simple. The relevant combination of probabilities is written in terms of marginals of the joint probability distribution, assumed to exist, arriving at the expression $\sum_{\alpha}c\naw{\alpha}p\naw{\alpha}$, where $c\naw{\alpha}$ are integers equal to the number of times $p\naw{\alpha}$ appears in the sum. Due to $0\leq p\naw{\alpha}\leq 1$, $\sum\limits_{\alpha}p\naw{\alpha}=1$ one obtains
\begin{equation}
\min\limits_{\alpha} c\naw{\alpha}\leq\sum_{\alpha}c\naw{\alpha}p\naw{\alpha}\leq \max\limits_{\alpha} c\naw{\alpha}.\label{a4}
\end{equation}
In order to get the standard form of Bell inequalities one should express $p\naw{\alpha}$ in terms of relevant correlation functions.

In order to establish the violation of Bell inequalities in quantum mechanics one has to construct the particular examples. In two papers mentioned above \cite{ugur}, \cite{ugur1} G\H uney and Hillery proposed to use the group theoretical methods. Consider some finite group $G$ and its irreducible representation $D$. The space carrying the representation $D$ becomes the space of states of one party. One selects an orbit $\poisson{D\naw{g}\ket{\varphi}}_{g\in G}$ in such a way that it decomposes into disjoint sets of orthonormal bases. These bases define the spectral decompositions of observables entering the example. The space of states of the second party carries the second representation in the product $D\otimes D$; the corresponding orbits read $\poisson{D\naw{g}\ket{\psi}}_{g\in G}$ and defines the observables of second party.\\
Let us construct the operator
\begin{equation}
X\naw{\varphi,\psi}\equiv\sum_{g\in G}\naw{D\naw{g}\ket{\varphi}\otimes D\naw{g}\ket{\psi}}\naw{\bra{\varphi}D^+\naw{g}\otimes\bra{\psi}D^+\naw{g}}.
\end{equation}
Defining
\begin{equation}
\begin{split}
& \ket{g,\varphi}\equiv D\naw{g}\ket{\varphi},\qquad \ket{g,\psi}\equiv D\naw{g}\ket{\psi}\\
& \ket{g,\varphi,\psi}\equiv \ket{g,\varphi}\otimes \ket{g,\psi}
\end{split}
\end{equation}
one finds for arbitrary bipartite state $\ket{\chi}$
\begin{equation}
\bra{\chi}X\ket{\chi}=\sum_{g\in G}\modu{\av{g,\varphi,\psi | \chi}}^2.\label{a3}
\end{equation}
The right hand side of eq. (\ref{a3}) represents the sum of probabilities of particular outcomes of measurement performed on observables defined by the orbits $\poisson{\ket{g,\varphi}}$ and $\poisson{\ket{g,\psi}}$. Its maximal value corresponds to maximal eigenvector of $X$. In this way we obtain a kind of Cirel'son bound \cite{cirelson} for the class of states under consideration.

On the other hand one easily derives the Bell inequality involving the sum of probabilities on the right hand side of eq. (\ref{a3}). To this end one assumes the existence of joint probability distribution for all observables defined by both orbits (note that the ones belonging to one orbit in general do not commute) and uses the inequalities (\ref{a4}).

It remains to find the maximal eigenvalue of $X$. To this end assume that in the decomposition of $D\otimes D$ into irreducible pieces, 
\begin{equation}
D\otimes D=\bigoplus\limits_{s}D^{\naw{s}}
\end{equation}
each $D^{\naw{s}}$ appears only once. Then, by Schur's lemma, $X\naw{\varphi,\psi}$ is diagonal and reduces to a multiple of unity on each irreducible component. Using the orthogonality relations it is easy to see that the relevant eigenvalues of $X\naw{\varphi,\psi}$ are \cite{ugur1}
\begin{equation}
\frac{\modu{G}}{d_s}\parallel\naw{\ket{\varphi}\otimes\ket{\psi}}_s\parallel^2\label{b1}
\end{equation}
where $\modu{G}$ is the order of $G$, $d_s$ - the dimension of $D^{\naw{s}}$ and $\naw{\ket{\varphi}\otimes\ket{\psi}}_s$ is the projection of $\ket{\varphi}\otimes \ket{\psi}$ on the carrier space of $D^{\naw{s}}$.

In general, in order to break the Bell inequality it is necessary to consider a number of orbits. To this end one considers the orbits generated by $N$ pairs of vectors $\naw{\ket{\varphi_n},\ket{\psi_n}}$ and the corresponding operators $X\naw{\varphi_n,\psi_n}$. They mutually commute so the eigenvalues of 
\begin{equation}
X=\sum_{n=1}^N X\naw{\varphi_n,\psi_n}\label{b4}
\end{equation}
are the sums of eigenvalues of all $X\naw{\varphi_n,\psi_n}$. In this way one can maximize the sum of probabilities
\begin{equation}
\sum_{n=1}^N\sum_{g\in G}\modu{\av{g,\varphi_n,\psi_n |\chi}}^2
\end{equation}
and proceed as above.

\section{The $S_4$ group: three orbits}

$S_4$ is the group of order 24. It has 6 conjugancy classes. There exist six irreducible representations of $S_4$: trivial representation, the alternating representation, the homomorphic twodimensional one and two threedimansional representations, $D$ and $\widetilde{D}$; $\widetilde{D}$ is obtained from $D$ by multiplication by the alternating representation. All representations can be made orthogonal. 

Consider the threedimensional representation $D$. It can be described by writing out the matrices representing the transpositions:
\begin{equation}
D\naw{12}=\left[\begin{array}{ccc}
1 & 0 & 0\\
0 & 1 & 0\\
0 & 0 & -1
\end{array}\right],\qquad
 D\naw{13}=\left[\begin{array}{ccc}
1 & 0 & 0\\
0 & -\frac{1}{2} & -\frac{\sqrt{3}}{2}\\
0 & -\frac{\sqrt{3}}{2} & \frac{1}{2}
\end{array}\right]
\end{equation} 
\begin{equation}
D\naw{14}=\left[\begin{array}{ccc}
-\frac{1}{3} & -\frac{\sqrt{2}}{3} & -\frac{\sqrt{6}}{3}\\
-\frac{\sqrt{2}}{3} & \frac{5}{6} & -\frac{\sqrt{3}}{6}\\
-\frac{\sqrt{6}}{3} & -\frac{\sqrt{3}}{6}& \frac{1}{2}
\end{array}\right],\qquad
 D\naw{23}=\left[\begin{array}{ccc}
1 & 0 & 0\\
0 & -\frac{1}{2} & \frac{\sqrt{3}}{2}\\
0 & \frac{\sqrt{3}}{2} & \frac{1}{2}
\end{array}\right]
\end{equation}
\begin{equation}
D\naw{24}=\left[\begin{array}{ccc}
-\frac{1}{3} & -\frac{\sqrt{2}}{3} & \frac{\sqrt{6}}{3}\\
-\frac{\sqrt{2}}{3} & \frac{5}{6} & \frac{\sqrt{3}}{6}\\
\frac{\sqrt{6}}{3} & \frac{\sqrt{3}}{6}& \frac{1}{2}
\end{array}\right],\qquad
 D\naw{34}=\left[\begin{array}{ccc}
-\frac{1}{3} & \frac{\sqrt{8}}{3} & 0\\
\frac{\sqrt{8}}{3} & \frac{1}{3} & 0\\
0 & 0 & 1
\end{array}\right].
\end{equation}
$S_4$ is the symmetry group of regular tetrahedron. One can make the correspondence between the symmetries of tetrahedron and the representation $D$. To this end we find an (degenerate) orbit which forms a tetrahedron. It is easy to check that the vectors:  $\vec{a}_1=\naw{-\frac{1}{3},-\frac{\sqrt{2}}{3},-\frac{\sqrt{6}}{3}}$,  $\vec{a}_2=\naw{-\frac{1}{3},-\frac{\sqrt{2}}{3},\frac{\sqrt{6}}{3}}$,  $\vec{a}_3=\naw{-\frac{1}{3},\frac{\sqrt{8}}{3},0}$  and $\vec{a}_4=\naw{1,0,0}$ form the orbit of $S_4$ and are the vertices of regular tetrahedron.

Now, let us select the orbits we will use in the construction of our examples of breaking the Bell inequalities. A generic orbit consists of 24 states. According to the discussion in Sec. II we should choose the orbit consisting of eight triples of orthonormal vectors, each providing the spectral decomposition of one of the eight observables. The elements of the orbit are numbered as $\ket{x_\alpha^i}$, $i=1,\ldots,8$, $\alpha=0,1,2$. Then we demand $\av{x_\alpha^i | x_\beta^i}=\delta_{\alpha\beta}$, $i=1,\ldots ,8$ (no summation over $i$). Consequently, we are dealing with twice the eight observables ($a_i$ stands for $"$Alice$"$ and $b_i$ for $"$Bob$"$) 
\begin{equation}
a_i=\sum_{\alpha=0}^2\alpha\ket{x_\alpha^i}\bra{x_\alpha^i},\qquad b_i=\sum_{\beta=0}^2\beta\ket{x_\beta^i}\bra{x_\beta^i}.
\end{equation}
Taking into account that any element of $S_4$ is a product of transpositions and each transposition is represented by a reflection in the symmetry plane orthogonal to the edge connecting the transposed verticles we find the relevant orbit by simple examination of terahedron geometry\\
\begin{footnotesize}
$a_1:\quad \naw{x_0^1,x_1^1,x_2^1}$
\begin{equation}
\ket{x_0^1}=\left[\begin{array}{c}
\frac{\sqrt{3}}{3}\\
\frac{\sqrt{3}}{3}\\
-\frac{\sqrt{3}}{3}
\end{array}\right],
\quad \ket{x_1^1}=\left[\begin{array}{c}
\frac{\sqrt{3}}{3}\\
\frac{1}{2}\naw{1-\frac{\sqrt{3}}{3}}\\
\frac{1}{2}\naw{1+\frac{\sqrt{3}}{3}}\\
\end{array}\right],\quad 
\ket{x_2^1}=\left[\begin{array}{c}
\frac{\sqrt{3}}{3}\\
-\frac{1}{2}\naw{1+\frac{\sqrt{3}}{3}}\\
-\frac{1}{2}\naw{1-\frac{\sqrt{3}}{3}}\\
\end{array}\right]\label{a5}
\end{equation}
$a_2:\quad \naw{x_0^2,x_1^2,x_2^2}$
\begin{equation}
\ket{x_0^2}=\left[\begin{array}{c}
\frac{1}{9}\naw{-3\sqrt{2}-\sqrt{3}-\sqrt{6}}\\
\frac{1}{18}\naw{-3+5\sqrt{3}-2\sqrt{6}}\\
\frac{1}{6}\naw{-1-2\sqrt{2}+\sqrt{3}}\\
\end{array}\right],\,
 \ket{x_1^2}=\left[\begin{array}{c}
\frac{1}{9}\naw{-\sqrt{3}+2\sqrt{6}}\\
\frac{1}{18}\naw{9-\sqrt{3}-2\sqrt{6}}\\
-\frac{1}{6}\naw{1+2\sqrt{2}+\sqrt{3}}\\
\end{array}\right],\,
\ket{x_2^2}=\left[\begin{array}{c}
\frac{1}{9}\naw{3\sqrt{2}-\sqrt{3}-\sqrt{6}}\\
-\frac{1}{9}\naw{3+2\sqrt{3}+\sqrt{6}}\\
\frac{1}{3}\naw{1-\sqrt{2}}\\
\end{array}\right],
\end{equation}
$a_3:\quad \naw{x_0^3,x_1^3,x_2^3}$
\begin{equation}
\ket{x_0^3}=\left[\begin{array}{c}
\frac{1}{9}\naw{3\sqrt{2}-\sqrt{3}-\sqrt{6}}\\
\frac{1}{18}\naw{3+5\sqrt{3}-2\sqrt{6}}\\
\frac{1}{6}\naw{1+2\sqrt{2}+\sqrt{3}}\\
\end{array}\right],\,
 \ket{x_1^3}=\left[\begin{array}{c}
\frac{1}{9}\naw{-\sqrt{3}+2\sqrt{6}}\\
-\frac{1}{18}\naw{9+\sqrt{3}+2\sqrt{6}}\\
\frac{1}{6}\naw{1+2\sqrt{2}-\sqrt{3}}\\
\end{array}\right],\,
\ket{x_2^3}=\left[\begin{array}{c}
-\frac{1}{9}\naw{3\sqrt{2}+\sqrt{3}+\sqrt{6}}\\
\frac{1}{9}\naw{3-2\sqrt{3}-\sqrt{6}}\\
\frac{1}{3}\naw{-1+\sqrt{2}}\\
\end{array}\right],
\end{equation}
$a_4:\quad \naw{x_0^4,x_1^4,x_2^4}$
\begin{equation}
\ket{x_0^4}=\left[\begin{array}{c}
\frac{1}{9}\naw{3\sqrt{2}-\sqrt{3}-\sqrt{6}}\\
\frac{1}{18}\naw{3-\sqrt{3}+4\sqrt{6}}\\
\frac{1}{6}\naw{3+\sqrt{3}}\\
\end{array}\right],\,
 \ket{x_1^4}=\left[\begin{array}{c}
\frac{1}{9}\naw{-\sqrt{3}+2\sqrt{6}}\\
\frac{1}{9}\naw{\sqrt{3}+2\sqrt{6}}\\
-\frac{\sqrt{3}}{3}
\end{array}\right],\,
\ket{x_2^4}=\left[\begin{array}{c}
-\frac{1}{9}\naw{3\sqrt{2}+\sqrt{3}+\sqrt{6}}\\
\frac{1}{18}\naw{-3-\sqrt{3}+4\sqrt{6}}\\
\frac{1}{6}\naw{-3+\sqrt{3}}\\
\end{array}\right],
\end{equation}
$a_5:\quad \naw{x_0^5,x_1^5,x_2^5}$
\begin{equation}
\ket{x_0^5}=\left[\begin{array}{c}
\frac{1}{9}\naw{-\sqrt{3}+2\sqrt{6}}\\
\frac{1}{18}\naw{9-\sqrt{3}-2\sqrt{6}}\\
\frac{1}{6}\naw{1+2\sqrt{2}+\sqrt{3}}\\
\end{array}\right],\,
 \ket{x_1^5}=\left[\begin{array}{c}
-\frac{1}{9}\naw{3\sqrt{2}+\sqrt{3}+\sqrt{6}}\\
\frac{1}{18}\naw{-3+5\sqrt{3}-2\sqrt{6}}\\
\frac{1}{6}\naw{1+2\sqrt{2}-\sqrt{3}}\\
\end{array}\right],\,
\ket{x_2^5}=\left[\begin{array}{c}
\frac{1}{9}\naw{3\sqrt{2}-\sqrt{3}-\sqrt{6}}\\
-\frac{1}{9}\naw{3+2\sqrt{3}+\sqrt{6}}\\
\frac{1}{3}\naw{-1+\sqrt{2}}\\
\end{array}\right],
\end{equation}
$a_6:\quad \naw{x_0^6,x_1^6,x_2^6}$
\begin{equation}
\ket{x_0^6}=\left[\begin{array}{c}
-\frac{1}{9}\naw{3\sqrt{2}+\sqrt{3}+\sqrt{6}}\\
\frac{1}{18}\naw{-3-\sqrt{3}+4\sqrt{6}}\\
\frac{1}{6}\naw{3-\sqrt{3}}
\end{array}\right],\,
 \ket{x_1^6}=\left[\begin{array}{c}
\frac{1}{9}\naw{3\sqrt{2}-\sqrt{3}-\sqrt{6}}\\
\frac{1}{18}\naw{3-\sqrt{3}+4\sqrt{6}}\\
-\frac{1}{6}\naw{3+\sqrt{3}}
\end{array}\right],\,
\ket{x_2^6}=\left[\begin{array}{c}
\frac{1}{9}\naw{-\sqrt{3}+2\sqrt{6}}\\
\frac{1}{9}\naw{\sqrt{3}+2\sqrt{6}}\\
\frac{\sqrt{3}}{3}
\end{array}\right],
\end{equation}
$a_7:\quad \naw{x_0^7,x_1^7,x_2^7}$
\begin{equation}
\ket{x_0^7}=\left[\begin{array}{c}
\frac{1}{9}\naw{3\sqrt{2}-\sqrt{3}-\sqrt{6}}\\
\frac{1}{18}\naw{3+5\sqrt{3}-2\sqrt{6}}\\
-\frac{1}{6}\naw{1+2\sqrt{2}+\sqrt{3}}
\end{array}\right],\,
 \ket{x_1^7}=\left[\begin{array}{c}
\frac{1}{9}\naw{-\sqrt{3}+2\sqrt{6}}\\
-\frac{1}{18}\naw{9+\sqrt{3}+2\sqrt{6}}\\
\frac{1}{6}\naw{-1-2\sqrt{2}+\sqrt{3}}\\
\end{array}\right],\,
\ket{x_2^7}=\left[\begin{array}{c}
-\frac{1}{9}\naw{3\sqrt{2}+\sqrt{3}+\sqrt{6}}\\
\frac{1}{9}\naw{3-2\sqrt{3}-\sqrt{6}}\\
\frac{1}{3}\naw{1-\sqrt{2}}\\
\end{array}\right],
\end{equation}
$a_8:\quad \naw{x_0^8,x_1^8,x_2^8}$
\begin{equation}
\ket{x_0^8}=\left[\begin{array}{c}
\frac{\sqrt{3}}{3}\\
-\frac{1}{2}\naw{1+\frac{\sqrt{3}}{3}}\\
\frac{1}{2}\naw{1-\frac{\sqrt{3}}{3}}
\end{array}\right],\,
 \ket{x_1^8}=\left[\begin{array}{c}
\frac{\sqrt{3}}{3}\\
\frac{1}{2}\naw{1-\frac{\sqrt{3}}{3}}\\
-\frac{1}{2}\naw{1+\frac{\sqrt{3}}{3}}
\end{array}\right],\,
\ket{x_2^8}=\left[\begin{array}{c}
\frac{\sqrt{3}}{3}\\
\frac{\sqrt{3}}{3}\\
\frac{\sqrt{3}}{3}
\end{array}\right].\label{a6}
\end{equation}
\end{footnotesize}
For all examples given below the states describing both parties beolng to the same orbit defined by eqs. (\ref{a5})$\div$(\ref{a6}). However, in each case the orbit of second party ($"$Bob$"$) will be shifted with respect to the one of first party ($"$Alice$"$). 

In order to compute the eigenvalues of the operator $X$ we should know the matrix which relates the product basis to the one in which the decomposition
\begin{equation}
D\otimes D=D\oplus \widetilde{D}\oplus D_2\oplus D_0\label{b3}
\end{equation}  
is explicit. It reads
\begin{equation}
C=\left [\begin{array}{ccccccccc}
\sqrt{\frac{2}{3}} & 0 & 0 & 0 & -\frac{1}{\sqrt{6}} & 0 & 0 & 0 & -\frac{1}{\sqrt{6}}\\
0 & -\frac{1}{\sqrt{6}} & 0 & -\frac{1}{\sqrt{6}} & \frac{1}{\sqrt{3}} & 0 & 0 & 0 & -\frac{1}{\sqrt{3}}\\
0 & 0 & -\frac{1}{\sqrt{6}} & 0 & 0 & -\frac{1}{\sqrt{3}} & -\frac{1}{\sqrt{6}} & -\frac{1}{\sqrt{3}} & 0\\
0 & \frac{1}{\sqrt{2}} & 0 & -\frac{1}{\sqrt{2}} & 0 & 0 & 0 & 0 & 0\\
0 & 0 & \frac{1}{\sqrt{2}} & 0 & 0 & 0 & -\frac{1}{\sqrt{2}} & 0 & 0\\
 0 & 0 & 0 & 0 & 0 & \frac{1}{\sqrt{2}} & 0 & -\frac{1}{\sqrt{2}} &  0\\
 0 & \frac{1}{\sqrt{3}} & 0 & \frac{1}{\sqrt{3}} & \frac{1}{\sqrt{6}} & 0 & 0 & 0 & -\frac{1}{\sqrt{6}}\\
 0 & 0 & \frac{1}{\sqrt{3}} & 0 & 0 & -\frac{1}{\sqrt{6}} & \frac{1}{\sqrt{3}} & -\frac{1}{\sqrt{6}} & 0\\
 \frac{1}{\sqrt{3}} & 0 & 0 & 0 & \frac{1}{\sqrt{3}} & 0 & 0 & 0 & \frac{1}{\sqrt{3}}
\end{array}\right].\label{b2}
\end{equation}
Eqs. (\ref{b1}) and (\ref{b2}) allow us to compute all eigenvalues of arbitrary operator $X\naw{\varphi,\psi}$. In all examples given below the largest eigenvalue $\lambda_{max}$ correspond to the scalar component in the decomposition (\ref{b3}). All of them involve three orbits (N=3 in eq. (\ref{b4})).
\newpage
\underline{Example I}:
\begin{itemize}
\item First orbit\\
$\ket{\varphi_1}=\ket{x_0^1},\quad \ket{\psi_1}=\ket{x_1^4}$\\
$\lambda_{max}\simeq 7,40$
\item Second orbit\\
$\ket{\varphi_2}=\ket{x_0^1},\quad \ket{\psi_2}=\ket{x_0^7}$\\
$\lambda_{max}\simeq 4,57$
\item Third orbit\\
$\ket{\varphi_3}=\ket{x_0^1},\quad \ket{\psi_3}=\ket{x_1^5}$\\
$\lambda_{max}\simeq 4,12$
\end{itemize}
The maximal eigenvalue of the operator $X$ defined by eq. (\ref{b4}):\\
$\lambda_{max}(X)\simeq 16,09$.\\

\underline{Example II}:
\begin{itemize}
\item First orbit\\
$\ket{\varphi_1}=\ket{x_0^1},\quad \ket{\psi_1}=\ket{x_2^3}$\\
$\lambda_{max}\simeq 5,21$
\item Second orbit\\
$\ket{\varphi_2}=\ket{x_0^1},\quad \ket{\psi_2}=\ket{x_1^6}$\\
$\lambda_{max}\simeq 5,30$
\item Third orbit\\
$\ket{\varphi_3}=\ket{x_0^1},\quad \ket{\psi_3}=\ket{x_0^1}$\\
$\lambda_{max}\simeq 8,00$
\end{itemize}
The maximal eigenvalue of $X$:\\
$\lambda_{max}(X)\simeq 18,51$.
\newpage
\underline{Example III}:
\begin{itemize}
\item First orbit\\
$\ket{\varphi_1}=\ket{x_0^1},\quad \ket{\psi_1}=\ket{x_2^5}$\\
$\lambda_{max}\simeq 3,35$
\item Second orbit\\
$\ket{\varphi_2}=\ket{x_0^1},\quad \ket{\psi_2}=\ket{x_1^4}$\\
$\lambda_{max}\simeq 7,40$
\item Third orbit\\
$\ket{\varphi_3}=\ket{x_0^1},\quad \ket{\psi_3}=\ket{x_1^8}$\\
$\lambda_{max}\simeq 6,63$
\end{itemize}
The maximal eigenvalue of $X$:\\
$\lambda_{max}(X)\simeq 17,38$.\\
The corresponding sums of probabilities appearing on the right hand side of eq. (\ref{a3}) are written out explicitly in Appendix. Having computed the (maximal) quantum mechanical values of the relevant sums of probabilities one can study the corresponding Bell inequalities. To this end we compute the coefficients $c(\alpha)$ entering the inequalities (\ref{a4}). There are 16 observables 8 for $"$Alice$"$ and 8 for $"$Bob$"$. Therefore, the assumed joint probability is defined for $3^{16}$ configurations. We used the computer to check, for three examples above, how many times any given configuration appears in 72 terms in $"$classical$"$ counterpart of the right hand side of eq. (\ref{a3}). The result are summarized in Appendix. It follows that the relevant sums of probabilities have the upper bounds 16, 18 and 16 for the examples I, II and III, respectively. This implies that in all three examples the Bell inequalities are broken. 

\section{Interpretation in terms of game theory}
As it has been described in Refs. \cite{ugur} and \cite{ugur1} the Bell inequalities can be discussed in terms of a nonlocal game. To this end we assume there are two players, Alice and Bob and an arbitrator who sends Alice a value $s$ and Bob a value $t$, $s=1,2,\ldots,8$, $t=1,2\ldots,8$; assume that all of 64 possible values of $\naw{s,t}$ are equally likely. After receiving the numbers $s$ and $t$ from an arbitrator both Alice nad Bob transmit back the numbers $a$ and $b$, respectively, where $a=0,1,2$, $b=0,1,2$. They win iff the configuration $\naw{a_s=a,b_t=b}$ appears in the sum of probabilities corresponding to the right hand side of eq. (\ref{a3}). Let us consider for definiteness the example I. Using (\ref{d}) we get the set of winning values given in Table 1.

\begin{table}[t]
\caption{Winning configurations for nonlocal game defined by three orbits from example I}
\center
\small
\begin{tabular}{|l|l|}
\hline
s, t & Alice, Bob \\
\hline
14 & 01, 10, 22\\
15 & 01, 10, 22\\
17 & 00, 12, 21\\
24 & 02, 11, 20\\
25 & 01, 10, 22\\
28 & 02, 11, 20\\
34 & 00, 11, 22\\
37 & 00, 11, 22\\
38 & 02, 10, 21\\
41 & 01, 10, 22\\
42 & 02, 11, 20\\
43 & 00, 11, 22\\
\hline
\end{tabular}
\begin{tabular}{|l|l|}
\hline
s, t & Alice, Bob \\
\hline
51 & 01, 10, 22\\
52 & 01, 10, 22\\
56 & 02, 10, 21\\
65 & 01, 12, 20\\
67 & 02, 10, 21\\
68 & 00, 11, 22\\
71 & 00, 12, 21\\
73 & 00, 11, 22\\
76 & 01, 12, 20\\
82 & 02, 11, 20\\
83 & 01, 12, 20\\
86 & 00, 11, 22\\
\hline
\end{tabular}
\end{table}

Following Ref. \cite{ugur1} we can show that the maximal classical probability of winning the game is determined by Bell inequality. In fact, let $f_A\naw{s}$ and $f_B\naw{t}$ be the strategies of Alice and Bob, respectively; the function $f_{A,B}$ take their values in the set $\poisson{0,1,2}$. Let $F\naw{a,b;s,t}$ be the characteristic function for the set of winning strategies. Then the winning probability for the given strategies $f_A$, $f_B$ is 
\begin{equation}
\frac{1}{64}\sum_{a,b=0}^2\sum_{s,t=1}^8 F\naw{a,b;s,t}\delta_{a,f_A\naw{s}}\delta_{b,f_B\naw{t}}.\label{c}
\end{equation}
Now, the sum entering the left hand side of Bell inequality can be written as 
\begin{equation}
\sum_{a,b=0}^2\sum_{s,t=1}^8 F\naw{a,b;s,t}p\naw{a_s=a, b_t=b}
\end{equation}
which is bounded, in example I, by 16 provided $p\naw{a_s=a,b_t=b}$ can be derived from a joint probability distribution. However, defining
\begin{equation}
p\naw{a_1,\dots,a_8,b_1,\ldots,b_8}\equiv \prod_{k=1}^8 \delta_{a_k, f_A(k)}\delta_{b_k, f_B(k)}
\end{equation}
we find that $p\naw{a_s,b_t}$ are derived as marginals from the above joint probability. Therefore, the success probability for any classical strategy $\naw{f_A(s),f_B(t)}$ cannot exceed $\frac{16}{64}=0,25$.\\
Note that the optimal strategy saturating this limit always exists. To see this let $\alpha=\naw{\underline{a}_1,\ldots,\underline{a}_8,\underline{b}_1,\ldots,\underline{b}_8}$ be one of configurations for which $c(\alpha)$ attains its maximal value. Then the Bell inequality is saturated for the joint distribution probability $p(\alpha)=1$, $p(\alpha')=0$ for $\alpha '\neq\alpha$. Such distribution can be written in form (\ref{c}) with $f_A(s)=\underline{a}_s$, $f_B(t)=\underline{b}_t$.

In the quantum strategy Alice and Bob share the state corresponding to the maximal eigenvalue of $\sum_{n=1}^3 X\naw{\varphi_n,\psi_n}$. If they receive the numbers $s$, $t$ from an arbitrator, they measure $a_s$ (Alice) and $b_t$ (Bob), respectively, and send the result to the arbitrator. The probability of winning in example I is then $\frac{16,09}{64}\simeq 0,2514$  which exceeds (although only slightly) the classical bound. Other examples can be treated similarly.

\begin{appendices}
\section{}
To make the results slightly more transparent we write out explicitly the sum of probabilities appearing on the right hand side of eq. (\ref{a3}). They read:

Example I
\begin{footnotesize}
\begin{equation}
\begin{split}
& S_1\equiv P\naw{a_1=0,b_4=1}+P\naw{a_1=1,b_5=0}+P\naw{a_1=2,b_7=1}
+P\naw{a_2=0,b_4=2}+\\
& \quad +P\naw{a_2=1,b_8=1}+P\naw{a_2=2,b_5=2}+P\naw{a_3=0,b_4=0}+P\naw{a_3=1,b_8=0}+\\
& \quad +P\naw{a_3=2,b_7=2}+P\naw{a_4=0,b_3=0}+P\naw{a_4=1,b_1=0}+P\naw{a_4=2,b_2=0}+\\
& \quad +P\naw{a_5=0,b_1=1}+P\naw{a_5=1,b_6=0}+P\naw{a_5=2,b_2=2}+P\naw{a_6=0,b_5=1}+\\
& \quad +P\naw{a_6=1,b_7=0}+P\naw{a_6=2,b_8=2}+P\naw{a_7=0,b_6=1}+P\naw{a_7=1,b_1=2}+\\
& \quad +P\naw{a_7=2,b_3=2}+P\naw{a_8=0,b_3=1}+P\naw{a_8=1,b_2=1}+P\naw{a_8=2,b_6=2}+\\
& \quad +P\naw{a_1=0,b_7=0}+P\naw{a_1=1,b_4=0}+P\naw{a_1=2,b_5=2}
+P\naw{a_2=0,b_5=1}+\\
& \quad +P\naw{a_2=1,b_4=1}+P\naw{a_2=2,b_8=0}+P\naw{a_3=0,b_8=2}+P\naw{a_3=1,b_7=1}+\\
& \quad +P\naw{a_3=2,b_4=2}+P\naw{a_4=0,b_1=1}+P\naw{a_4=1,b_2=1}+P\naw{a_4=2,b_3=2}+\\
& \quad +P\naw{a_5=0,b_6=2}+P\naw{a_5=1,b_2=0}+P\naw{a_5=2,b_1=2}+P\naw{a_6=0,b_7=2}+\\
& \quad +P\naw{a_6=1,b_8=1}+P\naw{a_6=2,b_5=0}+P\naw{a_7=0,b_1=0}+P\naw{a_7=1,b_3=1}+\\
& \quad +P\naw{a_7=2,b_6=0}+P\naw{a_8=0,b_2=2}+P\naw{a_8=1,b_6=1}+P\naw{a_8=2,b_3=0}+\\
& \quad +P\naw{a_1=0,b_5=1}+P\naw{a_1=1,b_7=2}+P\naw{a_1=2,b_4=2}
+P\naw{a_2=0,b_8=2}+\\
& \quad +P\naw{a_2=1,b_5=0}+P\naw{a_2=2,b_4=0}+P\naw{a_3=0,b_7=0}+P\naw{a_3=1,b_4=1}+\\
& \quad +P\naw{a_3=2,b_8=1}+P\naw{a_4=0,b_2=2}+P\naw{a_4=1,b_3=1}+P\naw{a_4=2,b_1=2}+\\
& \quad +P\naw{a_5=0,b_2=1}+P\naw{a_5=1,b_1=0}+P\naw{a_5=2,b_6=1}+P\naw{a_6=0,b_8=0}+\\
& \quad +P\naw{a_6=1,b_5=2}+P\naw{a_6=2,b_7=1}+P\naw{a_7=0,b_3=0}+P\naw{a_7=1,b_6=2}+\\
& \quad +P\naw{a_7=2,b_1=1}+P\naw{a_8=0,b_6=0}+P\naw{a_8=1,b_3=2}+P\naw{a_8=2,b_2=0}
\end{split}\label{d}
\end{equation}
\end{footnotesize}

Example II:
\begin{footnotesize}
\begin{equation}
\begin{split}
& S_2\equiv P\naw{a_1=0,b_3=2}+P\naw{a_1=1,b_2=0}+P\naw{a_1=2,b_6=0}
+P\naw{a_2=0,b_1=1}+\\
& \quad +P\naw{a_2=1,b_3=0}+P\naw{a_2=2,b_6=2}+P\naw{a_3=0,b_2=1}+P\naw{a_3=1,b_6=1}+\\
& \quad +P\naw{a_3=2,b_1=0}+P\naw{a_4=0,b_7=1}+P\naw{a_4=1,b_5=2}+P\naw{a_4=2,b_8=0}+\\
& \quad +P\naw{a_5=0,b_7=0}+P\naw{a_5=1,b_8=1}+P\naw{a_5=2,b_4=1}+P\naw{a_6=0,b_1=2}+\\
& \quad +P\naw{a_6=1,b_3=1}+P\naw{a_6=2,b_2=2}+P\naw{a_7=0,b_5=0}+P\naw{a_7=1,b_4=0}+\\
& \quad +P\naw{a_7=2,b_8=2}+P\naw{a_8=0,b_4=2}+P\naw{a_8=1,b_5=1}+P\naw{a_8=2,b_7=2}+\\
& \quad +P\naw{a_1=0,b_6=1}+P\naw{a_1=1,b_3=0}+P\naw{a_1=2,b_2=2}
+P\naw{a_2=0,b_6=0}+\\
& \quad +P\naw{a_2=1,b_1=0}+P\naw{a_2=2,b_3=1}+P\naw{a_3=0,b_6=2}+P\naw{a_3=1,b_1=2}+\\
& \quad +P\naw{a_3=2,b_2=0}+P\naw{a_4=0,b_5=0}+P\naw{a_4=1,b_8=1}+P\naw{a_4=2,b_7=2}+\\
& \quad +P\naw{a_5=0,b_8=2}+P\naw{a_5=1,b_4=2}+P\naw{a_5=2,b_7=1}+P\naw{a_6=0,b_3=2}+\\
& \quad +P\naw{a_6=1,b_2=1}+P\naw{a_6=2,b_1=1}+P\naw{a_7=0,b_4=1}+P\naw{a_7=1,b_8=0}+\\
& \quad +P\naw{a_7=2,b_5=1}+P\naw{a_8=0,b_5=2}+P\naw{a_8=1,b_7=0}+P\naw{a_8=2,b_4=0}+\\
& \quad +P\naw{a_1=0,b_1=0}+P\naw{a_1=1,b_1=1}+P\naw{a_1=2,b_1=2}
+P\naw{a_2=0,b_2=0}+\\
& \quad +P\naw{a_2=1,b_2=1}+P\naw{a_2=2,b_2=2}+P\naw{a_3=0,b_3=0}+P\naw{a_3=1,b_3=1}+\\
& \quad +P\naw{a_3=2,b_3=2}+P\naw{a_4=0,b_4=0}+P\naw{a_4=1,b_4=1}+P\naw{a_4=2,b_4=2}+\\
& \quad +P\naw{a_5=0,b_5=0}+P\naw{a_5=1,b_5=1}+P\naw{a_5=2,b_5=2}+P\naw{a_6=0,b_6=0}+\\
& \quad +P\naw{a_6=1,b_6=1}+P\naw{a_6=2,b_6=2}+P\naw{a_7=0,b_7=0}+P\naw{a_7=1,b_7=1}+\\
& \quad +P\naw{a_7=2,b_7=2}+P\naw{a_8=0,b_8=0}+P\naw{a_8=1,b_8=1}+P\naw{a_8=2,b_8=2}
\end{split}
\end{equation}
\end{footnotesize}

Example III:
\begin{footnotesize}
\begin{equation}
\begin{split}
& S_3\equiv P\naw{a_1=0,b_5=2}+P\naw{a_1=1,b_7=0}+P\naw{a_1=2,b_4=0}
+P\naw{a_2=0,b_8=0}+\\
& \quad +P\naw{a_2=1,b_5=1}+P\naw{a_2=2,b_4=1}+P\naw{a_3=0,b_7=1}+P\naw{a_3=1,b_4=2}+\\
& \quad +P\naw{a_3=2,b_8=2}+P\naw{a_4=0,b_2=1}+P\naw{a_4=1,b_3=2}+P\naw{a_4=2,b_1=1}+\\
& \quad +P\naw{a_5=0,b_2=0}+P\naw{a_5=1,b_1=2}+P\naw{a_5=2,b_6=2}+P\naw{a_6=0,b_8=1}+\\
& \quad +P\naw{a_6=1,b_5=0}+P\naw{a_6=2,b_7=2}+P\naw{a_7=0,b_3=1}+P\naw{a_7=1,b_6=0}+\\
& \quad +P\naw{a_7=2,b_1=0}+P\naw{a_8=0,b_6=1}+P\naw{a_8=1,b_3=0}+P\naw{a_8=2,b_2=2}+\\
& \quad +P\naw{a_1=0,b_4=1}+P\naw{a_1=1,b_5=0}+P\naw{a_1=2,b_7=1}
+P\naw{a_2=0,b_4=2}+\\
& \quad +P\naw{a_2=1,b_8=1}+P\naw{a_2=2,b_5=2}+P\naw{a_3=0,b_4=0}+P\naw{a_3=1,b_8=0}+\\
& \quad +P\naw{a_3=2,b_7=2}+P\naw{a_4=0,b_3=0}+P\naw{a_4=1,b_1=0}+P\naw{a_4=2,b_2=0}+\\
& \quad +P\naw{a_5=0,b_1=1}+P\naw{a_5=1,b_6=0}+P\naw{a_5=2,b_2=2}+P\naw{a_6=0,b_5=1}+\\
& \quad +P\naw{a_6=1,b_7=0}+P\naw{a_6=2,b_8=2}+P\naw{a_7=0,b_6=1}+P\naw{a_7=1,b_1=2}+\\
& \quad +P\naw{a_7=2,b_3=2}+P\naw{a_8=0,b_3=1}+P\naw{a_8=1,b_2=1}+P\naw{a_8=2,b_6=2}+\\
& \quad +P\naw{a_1=0,b_8=1}+P\naw{a_1=1,b_8=2}+P\naw{a_1=2,b_8=0}
+P\naw{a_2=0,b_7=2}+\\
& \quad +P\naw{a_2=1,b_7=0}+P\naw{a_2=2,b_7=1}+P\naw{a_3=0,b_5=0}+P\naw{a_3=1,b_5=2}+\\
& \quad +P\naw{a_3=2,b_5=1}+P\naw{a_4=0,b_6=2}+P\naw{a_4=1,b_6=1}+P\naw{a_4=2,b_6=0}+\\
& \quad +P\naw{a_5=0,b_3=0}+P\naw{a_5=1,b_3=2}+P\naw{a_5=2,b_3=1}+P\naw{a_6=0,b_4=2}+\\
& \quad +P\naw{a_6=1,b_4=1}+P\naw{a_6=2,b_4=0}+P\naw{a_7=0,b_2=1}+P\naw{a_7=1,b_2=2}+\\
& \quad +P\naw{a_7=2,b_2=0}+P\naw{a_8=0,b_1=2}+P\naw{a_8=1,b_1=0}+P\naw{a_8=2,b_1=1}
\end{split}
\end{equation}
\end{footnotesize}
Therefore, the corresponding Bell inequalities take the form
\begin{equation}
S_1\leq 16
\end{equation}
\begin{equation}
S_2\leq 18
\end{equation}
\begin{equation}
S_3\leq 16.
\end{equation}
They were obtained by assuming the existence of joint of random variables  $a_1,\ldots,a_8,$ $b_1,\ldots,b_8$ and computing the coefficients $c(\alpha)$ defined in Sec. II. More precisely, for each example we write all probabilities entering the sums $S_1$, $S_2$, $S_3$ as the marginals of joint probability distribution. As a result we obtain the expressions of the form 
\begin{equation}
S=\sum_\alpha c(\alpha)p(\alpha)
\end{equation}
where $\alpha$ runs over all $3^{16}$ configurations of the variables $a_1,\ldots,a_8,b_1,\ldots,b_8$. The results of numerical computations are summarized in the Table below.
\begin{table}[h]
\center
\begin{tabular}{|l|r|l|r|l|r|}
\hline
\multicolumn{2}{|c|}{Example I} & \multicolumn{2}{|c|}{Example II} & \multicolumn{2}{|c|}{Example III}\\
\hline
$c(\alpha)$ & No. of configurations & $c(\alpha)$ & No. of configurations &$c(\alpha)$ & No. of configurations \\
\hline
1 & 12 960 & 1 & 9 720 & 1 & 18 360\\
2 & 159 408 & 2 & 126 576 & 2 & 115 596\\
3 & 645 408 & 3 & 510 480 & 3 & 474 696\\
4 & 1 729 188 & 4 & 1 514 862 & 4 & 1 445 778\\
5 & 3 479 760 & 5 & 3 182 904 & 5 & 3 286 224\\
6 & 5 424 408 & 6 & 5 374584 & 6 & 5 510 160\\
7 & 6 896 016 & 7 & 7 139 664 & 7 & 7 178 976\\
8 & 7 261 569 & 8 & 7 822 791 & 8 & 7 670 547 \\
9 & 6 410 016 & 9 & 6 903 648 & 9 & 6 795 936\\
10 & 4 866 480 & 10 & 5 058 216 & 10 & 5 012 208\\
11 & 3 176 496 & 11 & 3 006 000 & 11 & 3 087 504\\
12 & 1 758 348 & 12 & 1 506 186 & 12 & 1 567 458\\
13 & 808 704 & 13 & 613 800 & 13 & 638 280\\
14 & 311 040 & 14 & 208 008 & 14 & 196 812\\
15 & 90 720 & 15 & 55 584 & 15 & 41 400\\
16 & 15 876 & 16 & 11 673 & 16 & 4 761\\
17 & 0 & 17 & 1 656 & 17 & 0\\
18 & 0 & 18 & 144 & 18 & 0\\
19 & 0 & 19 & 0 & 19 & 0\\
20 & 0 & 20 & 0 & 20 & 0\\
\hline
\end{tabular}
\end{table}
\end{appendices}

\subsection*{Acknowledgement}
Katarzyna Bolonek-Laso\'n would like to acknowledge Prof. Piotr Kosi\'nski  for helpful discussions and suggestions.  
Research of Katarzyna Bolonek-Laso\'n was supported by the NCN Grant no. DEC-2012/05/D/ST2/00754.

\end{document}